\documentclass[aps,prx,twocolumn,superscriptaddress]{revtex4-2}
\usepackage{graphicx}
\usepackage{dcolumn}
\usepackage{bm}
\usepackage{gensymb} 
\usepackage{amsmath,amssymb,bm,amsfonts,mathrsfs,epsfig,amsbsy,verbatim,bbm,mathtools,slashed} 
\usepackage{color} 
\usepackage[normalem]{ulem}

\makeatletter
\def\@fnsymbol#1{\ensuremath{\ifcase#1\or \dagger\or *\or \ddagger\or
   \mathsection\or \mathparagraph\or \|\or **\or \dagger\dagger
   \or \ddagger\ddagger \else\@ctrerr\fi}}
\makeatother

\begin{document}

\title{Anomalous Quantum Oscillations in Spin-3/2 Topological Semimetal YPtBi}

\author{Hyunsoo Kim}
\thanks{These two authors contributed equally.}
\affiliation{Maryland Quantum Materials Center, Department of Physics, University of Maryland, College Park, MD 20742, USA}

\author{Junhyun Lee}
\thanks{These two authors contributed equally.}
\affiliation{Department of Physics, Condensed Matter Theory Center and the Joint Quantum Institute, University of Maryland, College Park, Maryland 20742, USA}

\author{Halyna Hodovanets}
\affiliation{Maryland Quantum Materials Center, Department of Physics, University of Maryland, College Park, MD 20742, USA}

\author{Kefeng Wang}
\affiliation{Maryland Quantum Materials Center, Department of Physics, University of Maryland, College Park, MD 20742, USA}

\author{Jay D. Sau}
\affiliation{Department of Physics, Condensed Matter Theory Center and the Joint Quantum Institute, University of Maryland, College Park, Maryland 20742, USA}

\author{Johnpierre Paglione}
\email[]{paglione@umd.edu}
\affiliation{Maryland Quantum Materials Center, Department of Physics, University of Maryland, College Park, MD 20742, USA}
\affiliation{Canadian Institute for Advanced Research, Toronto, Ontario M5G 1Z8, Canada}

\date{\today}

\begin{abstract}
The proposed high-spin superconductivity in the half-Heusler compounds changes the landscape of superconductivity research. 
While superconducting instability is possible only in systems with quantum mechanically coherent quasiparticles, it has not been verified for any proposed high-spin Fermi surfaces.
Here we report an observation of anomalous Shubnikov-de Haas effect in half-Heusler YPtBi, which is compatible with a coherent $j=3/2$ Fermi surface. 
The quantum oscillation (QO) signal in cubic YPtBi manifests extreme anisotropy upon rotation of the magnetic field from [100] to [110] crystallographic direction where the QO signal drastically vanishes near [110]. 
This radical anisotropy for a cubic system cannot be explained by trivial scenarios for QO involving effective mass or impurity scattering, but it is naturally explained by the warping feature of the $j=3/2$ Fermi surface YPtBi.
Our results prove the high-spin nature of the quasiparticle in the half-Heusler compounds, which makes the realization of the unprecedented high-spin superconductivity more plausible.
\end{abstract}



\maketitle

{\it Introduction -} The intrinsic electron spin $s=1/2$ and its orbital angular momentum $l$ are often blended due to relativistic orbital motion.
This spin-orbit coupling (SOC) is very strong in compounds containing heavy elements, and therefore the total angular momentum, or effective spin $j$, becomes the most relevant quantum number \cite{Ho1999,Wu2003,Brydon2016,Yang2017,Kim2018}.
Changes in electronic band structure driven by SOC are fundamental to understanding non-trivial topology in quantum spin Hall effect \cite{Bernevig2006,Koenig2007} and Weyl physics \cite{Hirschberger2016,Suzuki2016}.
More recently, fermionic systems with $j$ greater than 1/2, stabilized by strong SOC, are gaining much attention because the possibility of high-spin quasiparticles in a solid-state opens up a new avenue for novel physics of interactions and resultant exotic phases of matter \cite{Brydon2016,Kim2018}.  
High-spin quasiparticles with $j=3/2$ can be found in the vicinity of quadratically touching bands in the well-known topological cubic materials such as pyrochlore iridates~\cite{Moon2013}, HgTe~\cite{Bernevig2006,Koenig2007}, or RPtBi half-Heuslers (R=rare earth)~\cite{Manna2018,Suzuki2016,Hirschberger2016}. Other $j=3/2$ systems include the cold-atom system~\cite{Wu2006,Kuzmenko2018}, anti-perovskite~\cite{Kawakami2018,Fang2020}, lacunar spinel~\cite{Jeong2017,Park2020}, and Rarita-Schwinger-Weyl semimetals~\cite{Boettcher2020}. 

Emergent phenomena stemming from the large $j$ are particularly interesting in the formation of Cooper pairs and superconducting states. 
Most notably, the pairing of high-spin fermions challenges the conventional picture of spin-$1/2$ superconductivity by disobeying the traditional singlet-triplet dichotomy~\cite{Ho1999}. 
High-spin pairing naturally allows additional large angular momentum Cooper pairs such as $J=2$ (quintet) and $J=3$ (septet) pairing states, and recent theoretical studies have unveiled these novel superconducting states are more stable than the conventional singlet/triplet states in large $j$ systems~\cite{Brydon2016,Kim2018}.

Depending on their symmetry, high-spin fermionic systems are predicted to host a number of distinct superconducting phases with unique properties~\cite{Venderbos2018}. 
In systems preserving both time reversal and inversion symmetries, a nematic $d$-wave can be imposed in the $s$-wave pairing channel in {\it cubic} compounds as the $d$-wave pairing causes spontaneous structural distortion~\cite{Boettcher2018,Sim2019}.
When time reversal symmetry is broken, the system favors a quintet Weyl superconductor with a charge neutral Bogoliubov Fermi surface as pseudo-magnetic field arises from the inter-band Cooper pairing~\cite{Agterberg2017,Timm2017}.
Unorthodox mixing between quintet $d$-wave and singlet $s$-wave is also expected even in a {\it centrosymmetric} superconductor~\cite{Yu2018,Yu2020}.
When inversion symmetry is broken, a singlet-septet pairing state with topological ring-shape line nodes can be realized, which manifests a 2D Majorana fluid enclosed by the surface projection of the nodal rings~\cite{Timm2017,Yang2017}. 
Hence high-spin superconductors serve as a potential shortcut to realizing a platform for fault-tolerant topological quantum computation.

\begin{figure}
\includegraphics[width=1\linewidth]{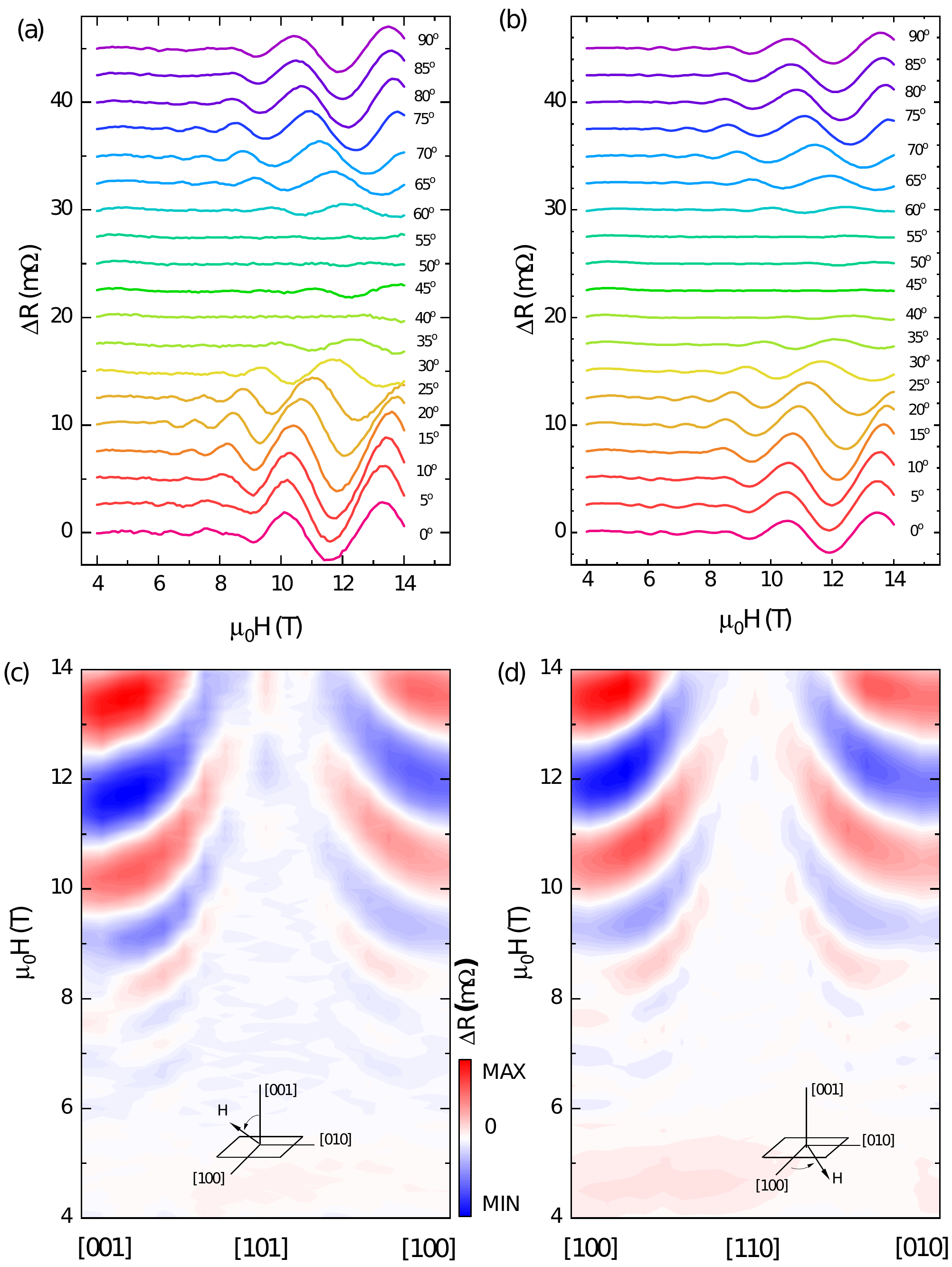}%
\caption{\label{fig1} Angle-dependent Shubnikov-de Haas quantum oscillations at $T=2$ K in YPtBi. The oscillatory components $\Delta R(T)$ are presented with various field-orientations (a) from [001] to [100] and (b) from [100] to [010]. Corresponding contour plots of $\Delta R$ are shown in (c) and (d), respectively, where the schematic of field rotation is shown. }
\end{figure}

The topological half-Heusler family RPtBi provides a perfect platform for hosting high-spin superconductivity, and thus the understanding of its topological band structure is of utmost importance. 
However, knowledge of the experimental band structure has been elusive as the bulk chemical potential is inconsistent between results from angle-resolved photoemission spectroscopy (ARPES)~\cite{Kim2018,Liu2011,Liu2016} and quantum oscillation (QO) experiments~\cite{Butch2011,Kim2018}. 
Furthermore, different interpretations of the observed surface states from different ARPES measurements make this issue more obscure~\cite{Kim2018,Liu2011,Liu2016}.
Rigorously verifying the Fermi surface in such systems and establishing its band structure is, therefore, an essential preliminary step. 
Moreover, further confirmation of the quasiparticle's coherence will be the basis of any high-spin phenomena, including the $j=3/2$ superconductivity.

In this work, we report compelling evidence for a coherent $j=3/2$ Fermi surface in YPtBi via studies of the angle-dependent Shubnikov-de Haas (SdH) effect.
The strikingly anisotropic variation of the amplitude of quantum oscillations in this cubic compound is compatible with the Fermi surface of $j=3/2$ quasiparticles.
Furthermore, our results offer a thorough understanding of the $j=3/2$ fermiology in the family of RPtBi compounds, confirming their topological nature of the band structure \cite{Kim2018}.
Therefore, this work provides a cornerstone for the realization of the unprecedented high-spin superconductivity as well as for developing applications towards topological superconductivity and fault-tolerant quantum computation \cite{Kitaev2003,Sau2010}. 
We also emphasize that our work successfully demonstrates the coherent $j=3/2$ Fermi surface, which has remained elusive in other high-spin systems \cite{Moon2013,Boettcher2020,HSKim2014,Park2020}, including the hole-doped silicon and germanium semiconductors which have been studied thoroughly for decades. 

{\it Experiment -} To probe the $j=3/2$ Fermi surface, we performed a comprehensive study of SdH quantum oscillations in YPtBi, an archetypical half-Heusler compound void of any localized magnetic moments.
Single crystals of YPtBi were grown out of molten Bi via the high-temperature flux method \cite{Canfield1991,Butch2011,Kim2018}. 
Electrical resistance was measured by using a standard four-probe technique in a commercial cryostat equipped with a 14 T magnet. 
The electrical contacts on the samples were attached by silver epoxy. 
A single-axis rotator was used to change the orientation of the samples with respect to the direction of an applied magnetic field. The orientations of the crystallographic direction were determined by using single-crystal x-ray diffraction patterns~\cite{SM}.

{\it Results} - Fig.~\ref{fig1} shows the SdH effect in YPtBi with various configurations at $T=2$ K. 
Panel (a) presents the oscillatory part of magnetoresistance $\Delta R$ which was obtained by subtracting a smoothly varying background magnetoresistance in a sample prepared out of the (001)-plane. 
(The magnetoresistance data are presented in the Supplementary Material~\cite{SM}.)
In this experiment, the magnetic field was rotated in the plane from [001] ($\theta = 0\degree$) to [100]-direction ($\theta =90\degree$), to reveal a remarkable angle-dependent amplitude with an oscillation pattern evidently symmetric about $\theta = 45\degree$. 
The QO frequency ($F \approx 45$ T) does not seem to significantly depend on the angle, which is consistent with a nearly spherical Fermi surface~\cite{Kim2018}. 
Panel (b) shows similar results from in-plane rotation experiments with the field direction from [100] to [010], consistent with the cubic symmetry.

\begin{figure}
\includegraphics[width=1\linewidth]{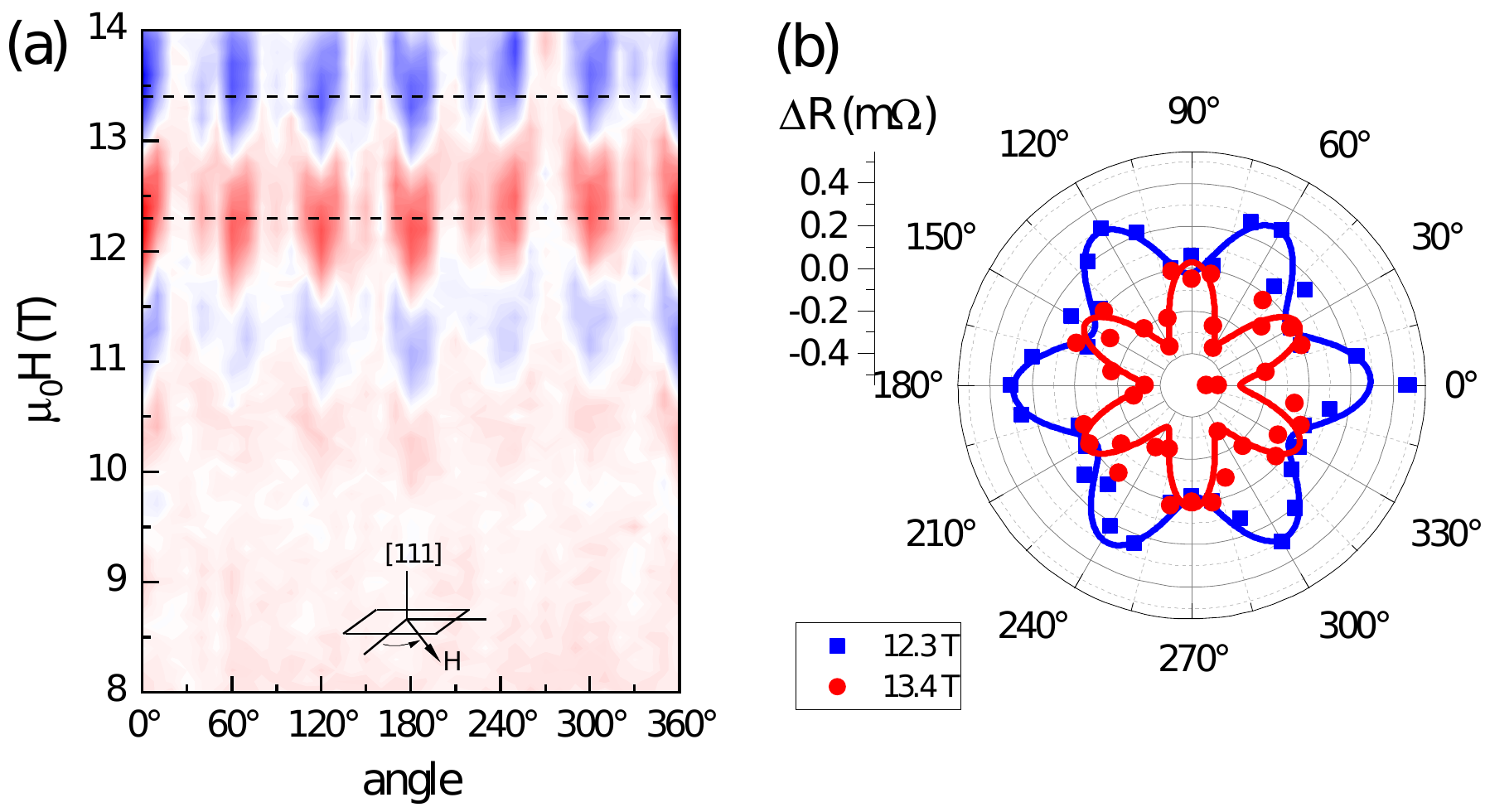}%
\caption{\label{fig2} Angle-dependent Shubnikov-de Haas quantum oscillations at $T=2$ K in YPtBi with magnetic fields rotating around [111]. (a) A contour plot of $\Delta R$ with the schematic of field configuration. (b) A polar plot of $\Delta R$ for magnetic fields of 12.3 T and 13.4 T (depicted in (a) with dashed lines). Both plots clearly show a six-fold symmetry.}
\end{figure}

Fig.~\ref{fig1}(c) and (d) display contour plots of $\Delta R(H,\theta)$ from these two rotation experiments.
We assigned the crystallographic orientations on the horizontal axis, according to the four-fold crystal symmetry of YPtBi.
The contour plots reveal a few key characteristics of the angle-dependent QO. 
Most notably, the amplitude of oscillations dramatically vanishes near the [110]-equivalent directions. 
Also, the oscillations move toward higher fields as approaching [110], and beating nodes were observed between $\theta = 0\degree$ and $\theta = 20\degree$ in the field rage around 7 T, indicating multiple oscillatory components. 

To confirm the vanishing QO amplitude along the [110] symmetry direction, a full-rotation experiment was performed on a sample cut out of the (111)-plane, with a magnetic field rotated in the sample plane. 
In this configuration, the field direction will come upon six [110]-equivalent directions, predicting a six-fold symmetry of $\Delta R(\theta)$. 
Fig.~\ref{fig2}(a) shows a contour plot of $\Delta R(H, \theta)$, which clearly shows a six-fold regular pattern.
Polar plot of $\Delta R(H,\theta)$ at $\mu_0H=13.4$ T and 12.3 T [Fig.\ref{fig2}(b)] shows that 
the angular amplitude of QO varies as $\cos(6\theta)$ (solid lines) for both cases, confirming the vanishing QO with the [110] symmetry directions.
The corresponding angle-dependent magnetoresistance data can be found in Supplementary Material~\cite{SM}.
Below we investigate the mechanism behind the observed drastic anisotropy in QO.

\begin{figure}
\includegraphics[width=1\linewidth]{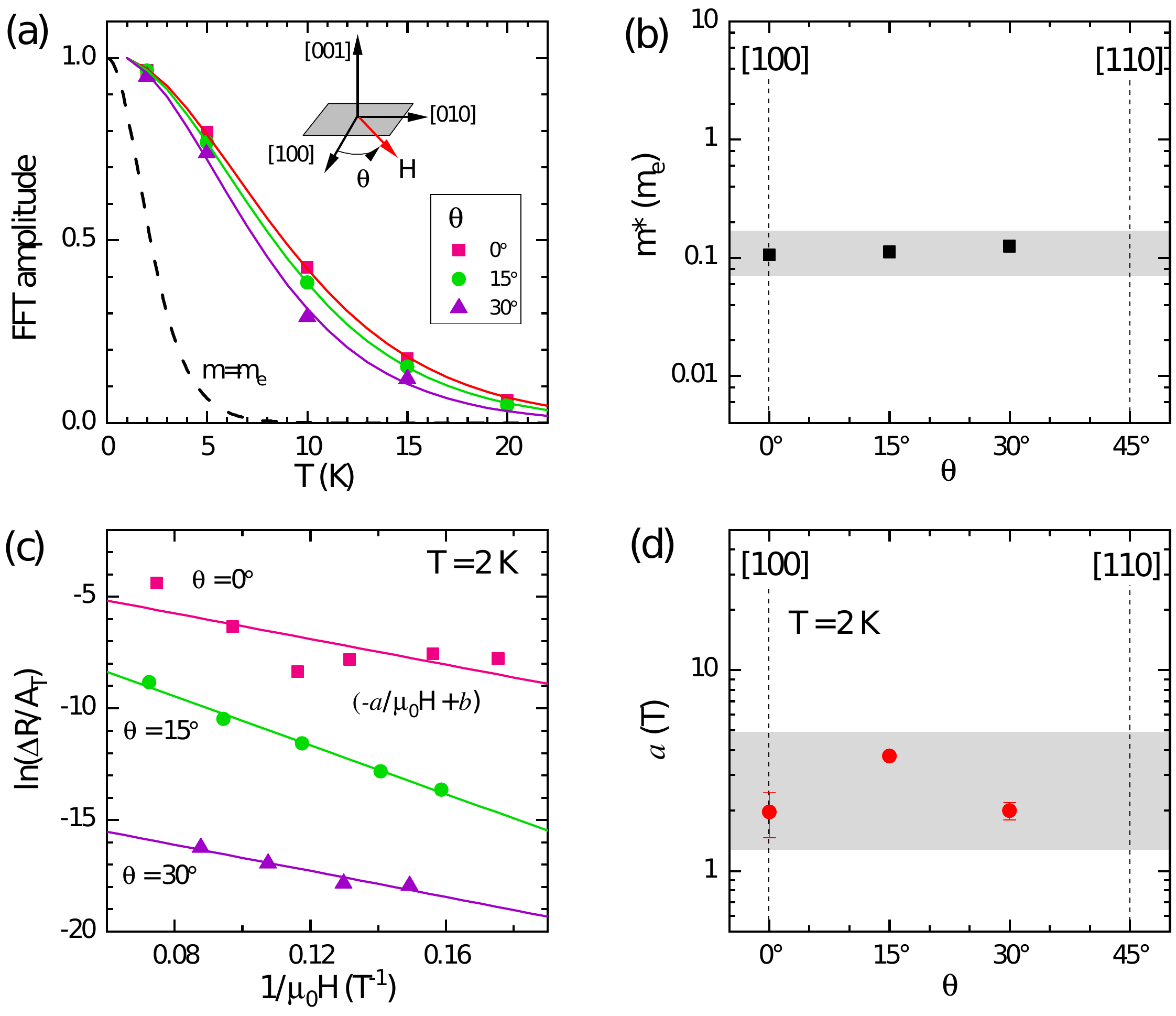}%
\caption{\label{fig3} Angle-dependent cyclotron mass $m^*$ and impurity scattering time $\tau$ in YPtBi. (a) Temperature dependence of the normalized QO amplitude. Symbols represent experimental values, and the solid lines represent the best theoretical fit with the LK formula $A_T(T,H)$ [Eq.~\eqref{eq:AT}]. (b) Angle-dependent $m^*$ obtained from the fit to Eq.~\eqref{eq:AT}. (c) Field-dependent QO amplitude at $T=2$ K. The symbols represent experimental values, and the solid straights lines represent a linear-fit $(-a/\mu_0H +b)$ to the LK formula $A_D(H)$ [Eq.~\eqref{eq:AD}]. (d) Angle-dependent slope $a$ of the linear fit in (c). }
\end{figure}

The QO amplitude is affected by the cyclotron mass $m^*$ and impurity scattering time $\tau$. 
In the semiclassical picture, QO is observable only when the cyclotron orbit can be completed, i.e., $\omega_c \tau <1$ where $\omega_c=e\mu_0 H/m^*$ is the cyclotron frequency. 
Therefore, the angle-dependence of $m^*$ and $\tau$ can potentially be responsible for the strong anisotropy in the observed QO amplitude. 

Within the standard Lifshitz-Kosevich (LK) theory~\cite{shoenberg1984}, the oscillatory part of the longitudinal magnetoresistance $\Delta R(T,H)$ is proportional to $A_T (T,H) A_D(H)$ where
\begin{align}\label{eq:AT}
    &A_T(T,H)=\frac{\alpha T /\mu_0 H}{\sinh(\alpha T /\mu_0 H)},\\
&A_D(H)= \exp{\left( -\frac{\alpha T_D }{\mu_0 H}\right)},\label{eq:AD}
\end{align}
with $\alpha=2\pi^2 k_B m^*/e\hbar$ and the Dingle temperature $T_D=\hbar/2\pi k_B \tau$. 
Evidently, $m^*$ and $\tau$ can be obtained from $T$-and $H$-dependence of the QO amplitude. 
However, $m^*$ and $\tau$ in the vicinity of [110]-direction have to be asymptotically deduced from the angular variation since QO is not observable in that orientation. 

Fig.~\ref{fig3}(a) shows the temperature evolutions of the QO amplitude, which are shown in the square, circle, and triangle symbols for $\theta = 0\degree$, 15\degree, and 30\degree, respectively. 
The QO amplitudes for various angles are extracted from the fast Fourier transform spectra~\cite{SM}.
The solid curves represent the fitting of Eq.~\eqref{eq:AT} to the data, from which the cyclotron mass $m^*$ can be determined. 
The obtained cyclotron mass is plotted in Fig.~\ref{fig3}(b), which shows little angle dependence.

The scattering time $\tau$ can be determined from the field variation of the QO amplitude at a given temperature. 
Fig.~\ref{fig3}(c) shows $\ln(\Delta R/A_T)$ vs. $1/\mu_0 H$ at $T=2$ K where the square, circle, and triangle symbols represent the data for $\theta = 0\degree$, 15\degree, and 30\degree, respectively.
The straight lines represent the best linear fit of $(-a/\mu_0H +b)$ to the data, where the slope is $a=\alpha T_D$ [Eq.~\eqref{eq:AD}]. 
The variation of $a$ which is proportional to $1/\tau$ only moderately depends on the angle as shown in Fig.~\ref{fig3}(d).

As we revealed in Fig.~\ref{fig3}(b) and (d), the observed angle-dependence of $m^*$ and $\tau$ in YPtBi have only marginal effect on the QO amplitude between [100] and [110]. 
Therefore, they do not account for the sudden vanishing of QO unless a nearly discontinuous change occurs between 30\degree~and 45\degree.
Although the diverging effective mass was observed in some unconventional superconductors in the vicinity of quantum critical point \cite{Ramshaw2015,Hashimoto2012}, this scenario is not plausible for YPtBi which is a low-carrier semimetal \cite{Canfield1991,Butch2011}. 

Apart from the $A_T (T,H) A_D(H)$ factor, the amplitude is also dependent on the density of states contributing to QO. 
In the LK formula~\cite{shoenberg1984}, this effect is included as a prefactor $[\partial^2_{k_\parallel} S (\tilde{k}_\parallel)]^{-1/2}$ of the amplitude of QO. 
Here, $S$ is the cross-section area of the Fermi surface perpendicular to $k_\parallel$, the momentum parallel to the external field, and $\tilde{k}_\parallel$ indicates the value of $k_\parallel$ where the Fermi surface area is an extrema~\footnote{Lifshitz-Kosevich formula only takes into account the second-order term of the curvature and thus has limitation in the case of YPtBi. However, the qualitative features persist and are sufficient to discuss the key features.}.
The strong angle-dependence of $[\partial^2_{k_\parallel} S (\tilde{k}_\parallel)]^{-1/2}$ could result in a drastic change in the QO amplitude upon rotation.
This effect has been well demonstrated in the systems with a corrugated 2D Fermi surface~\cite{Yamaji1989,Sebastian2014,Ramshaw2015}, but it has been overlooked in the 3D systems.

\begin{figure*}
\includegraphics[width=0.85\linewidth]{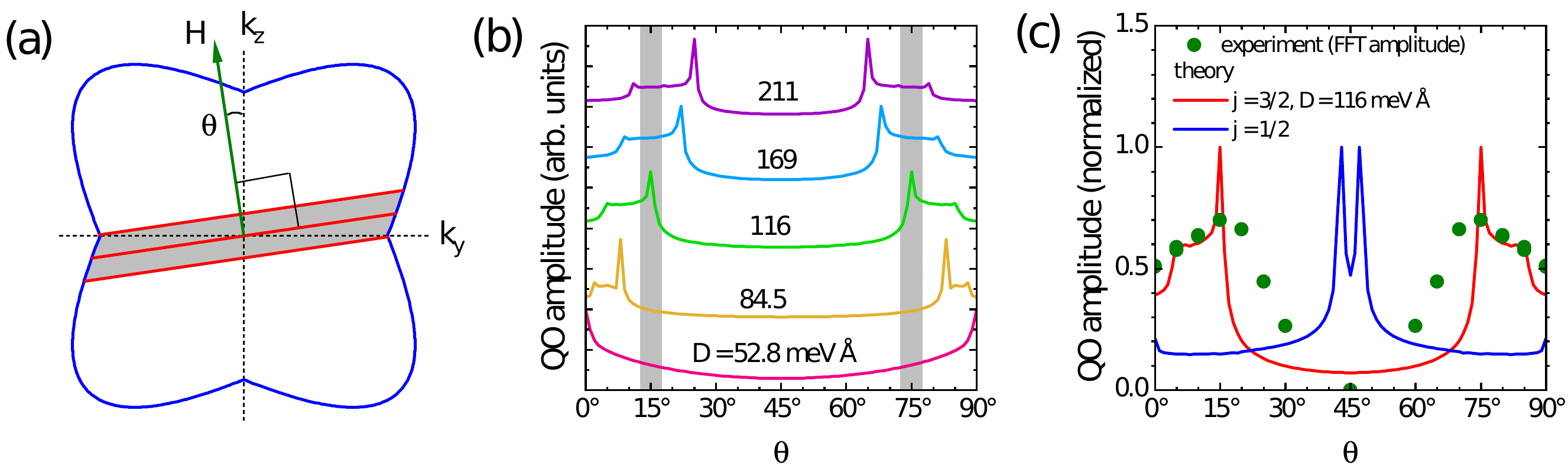}%
\caption{\label{fig4} Angular variation of the prefactor $[\partial^2_{k_\parallel} S (\tilde{k}_\parallel)]^{-1/2}$ of QO amplitude. (a) A schematic cross-section of $j=3/2$ Fermi surface, where the warping sensitively depends on $D$ [see Eq.~\eqref{eq:H0}]. The red lines and gray band represent the cyclotron orbits and the enhanced density of states on the Fermi surface perpendicular to the applied field $H$, when the quantum oscillation amplitude is maximum. (b) The angular variation of QO amplitude with different choices of $D$. The gray vertical bars represent angles for the maximum amplitude of experimental QO in YPtBi. (c) Angular variation of QO amplitude of $j=1/2$ and  $j=3/2$ Fermi surfaces compared to the experimental results (symbols). Although the determination of accurate angular variation requires full microscopic consideration, the relative position of the maximum and overall tendency show reasonable agreement with the experimental results. }
\end{figure*}

To determine the angular variation of $[\partial^2_{k_\parallel} S (\tilde{k}_\parallel)]^{-1/2}$, we construct the Fermi surface within the four-band $\bf k \cdot p$ model for spin $j=3/2$ electrons that is written as~\cite{Dresselhaus1955,Luttinger1956,Moon2013,Brydon2016,Cano2017}:
\begin{align}
	\mathcal{H}_0 =&A k^2 + B \sum_i k_i^2 J_i^2 + C \sum_{i \neq j}k_i k_j J_i J_j \nonumber \\
		&+ D \sum_i k_i (J_{i+1} J_i J_{i+1} + J_{i+2} J_i J_{i+2}) .
		\label{eq:H0}
\end{align}
Here, $J_i$'s are the $i$-th directional $j=3/2$ angular momentum operator, and we used $A = 22.9 \textrm{ eV} \textrm{\AA}^2$, $B = -20.7 \textrm { eV} \textrm{\AA}^2$, $C= -14.2 \textrm { eV} \textrm{\AA}^2$, and $D = 0.116 \textrm { eV} \textrm{\AA}$, which are previously determined in YPtBi~\footnote{The parameter values were adjusted based on the prior publications \cite{Brydon2016,Kim2018} to best describe the experimental SdH data.}.
The chemical potential of $\mu=-35$ meV, corresponding to the observed QO frequency $F=45$ T \cite{Butch2011,Kim2018}, is used in the following calculations.
Eq.~\eqref{eq:H0} gives a spin-split band structure, with only the principal axes being degenerate due to the $C_2$ rotational symmetries around the principal axes. 
The spin-split band structure with degenerate principal axes results in the bulging of the Fermi surface in the [111] direction. 
A schematic 2D projection of the warped Fermi surface of YPtBi with $k_z=0$ is depicted in Fig.~\ref{fig4}(a). 
One can find that $D$ in Eq. \eqref{eq:H0} is the dominant factor for the warping and therefore significantly affects the angular variation of $[\partial^2_{k_\parallel} S (\tilde{k}_\parallel)]^{-1/2}$. 

In Fig.~\ref{fig4}(b), we calculate the QO prefactor $[\partial^2_{k_\parallel} S (\tilde{k}_\parallel)]^{-1/2}$ of the outer Fermi surface as a function of $\theta$ for different values of $D$.
For $D\geq 84.5$ meV$\textrm{\AA}$, the prefactor clearly exhibits sharp peaks whose angular position depends on the choice of $D$. 
We note that the inner Fermi surface does not exhibit any sharp peak in the QO amplitude~\cite{SM}, and therefore the outer Fermi surface is likely responsible for the observed anisotropy in the QO amplitude.
We found that $[\partial^2_{k_\parallel} S (\tilde{k}_\parallel)]^{-1/2}$ exhibits the minimum value at $\theta=45\degree$ with all tested $D$ values, which suggests that the absence of QO around the [110]-equivalent directions stems from the intrinsic properties of the Fermi surface in YPtBi.

It is noteworthy that in YPtBi, apart from the vanishing QO in [110]-directions, the QO signal is strongest around $\theta = 15\degree$. We fine-tuned $D$ to match the experimental enhancement of QO and found that $D = 116$ meV$\textrm{\AA}$ best agrees with the experiment.
We plot the QO prefactor with  $D = 116$ meV$\textrm{\AA}$ together with the experimental data in Fig.~\ref{fig4}(c). 
Whereas the $j=3/2$ Fermi surface exhibits minimum near 45\degree, a hypothetical Fermi surface with $j=1/2$ produces {\it peaks} near the [110] direction, which is a result of the $j=1/2$ band structure having additional degenerate lines along [111]-direction~\cite{SM}.
Moreover, the angle-dependent Zeeman energy leads to a selective inter-orbit hopping during cyclotron motion, which additionally weakens the QO amplitude with a magnetic field near the [110] direction. 
This effect is discussed in the Supplementary Material~\cite{SM} in great detail.
Combining both effects, the angle-dependence of QO reasonably captures the $j=3/2$ nature of the Fermi surface in YPtBi.

{\it Discussion and summary -} In this work, we report an extreme amplitude variation of the quantum oscillations upon rotation of a magnetic field in the $j=3/2$ topological semimetal YPtBi. 
The quantum oscillation vanishes with an applied magnetic field along the [110]-equivalent crystallographic directions – completely at odds with the simple, nearly spherical band structure – and reaches a maximum at $\theta\approx 15\degree$ from the [100]-direction on the (001)-plane. 
We show that the observed quantum oscillations cannot be explained by angular variations of effective mass and impurity scattering, but can be naturally explained with the $j=3/2$ Fermi surface of YPtBi. 

The results of our work confirm the high-spin nature of the topological band structure in the half-Heusler RPtBi compounds.
Moreover, we experimentally demonstrate the existence of a coherent high-spin Fermi surface, which is the basis of the study of the novel spin-3/2 fermion system. 
We note that even in the cleanest doped-semiconductors such as Si and GaAs the Fermi surface of spin-3/2 hole-bands has not been observed due to inadequate mobility.
By laying this foundation for a new field of topological superconductivity, this work will not only change the landscape of superconductivity research but also establish a possible shortcut to the realization of fault-tolerant topological quantum devices.

{\it Acknowledgements - } The authors are grateful for useful discussions with D.~Agterberg, P.~Brydon, D.~Bulmash, P.~Li, E.-G.~Moon, A.~Nevidomskyy, V.~Yakovenko. This research was supported by the U.S. Department of Energy (DOE) Award No. DE-SC-0019154 (experimental investigations), and the Gordon and Betty Moore Foundation’s EPiQS Initiative through Grant No. GBMF9071 (materials synthesis).

\end{document}